\documentclass[twocolumn,english,american,twocolumn]{revtex4-2}
\usepackage[T1]{fontenc}
\usepackage[utf8]{inputenc}
\usepackage{amsmath}
\usepackage{amssymb}
\usepackage{graphicx}
\usepackage{babel}
\begin{document}
\title{Locking transition in coupled disordered systems}
\author{Guy Bunin}
\address{Department of Physics, Technion--Israel Institute of Technology,
Haifa 32000, Israel}
\begin{abstract}
When a system of many interacting constituents is extended across
space, every region carries the same disordered landscape, and it
is unclear whether distant regions settle into the same state. We
study copies of the Random Energy Model (REM), sharing one disorder
realization, coupled along a chain. A first-order transition separates
a phase with short-range correlations, from a locked phase in which
the entire chain occupies the lowest-energy state with probability
one, even at positive temperature. No intermediate ordering lengths
occur. In the Generalized REM the chain locks either to the lowest-energy
state or to the lowest free-energy valley.
\end{abstract}
\maketitle
Many systems of interest are built of numerous interacting constituents
whose concentrations may also vary in space. Cells and tissues contain
thousands of molecular species \citep{narayanankuttyMetabolicCoordinationPhase2025},
as do multi-component liquids of relevance to soft-matter \citep{searInstabilitiesComplexMixtures2003,shrinivasPhaseSeparation2021,jacobsSelfAssemblyBiomolecularCondensates2021},
and natural environments harbor individuals of many species \citep{royComplexInteractionsCan2020,pearceStabilizationExtensiveFinescale2020,buninDirectionalityCommunitylevelSelection2021,al-hiyasatSpatiotemporalNoiseStabilizes2026}.
In each case, the system may be viewed as many interacting spatially-varying
fields, or alternatively, discretizing space, as copies of the same
complex system coupled in space, e.g., by diffusion of the constituents.

In all these settings the interactions are numerous, specific and
often individually unknown. A standard, powerful approach is to treat
them as randomly drawn disorder, which brings the machinery of disordered
systems to bear \citep{wignerRandomMatricesPhysics1967,mayWillLargeComplex1972}.
When the systems are extended across space, each region of space carries
many variables, but since interactions are properties of the constituents
rather than position, every region carries the \emph{same} disorder.
So, for example, an energetically favorable state in one region is
favorable in every other. This is in contrast to systems such as random
magnets where disorder varies in space.

This class of systems is distinguished in that the number of variables
$N$ at every spatial position is large, so that the number of local
states, $2^{N}$, is exponentially large. This allows for behavior
with no analog when each position has only a few states, of which
the one-dimensional ordering studied here is one example. A long one-dimensional
chain of systems with finitely many states cannot order at equilibrium
at any positive temperature, since domain walls appear with finite
probability per unit length and destroy long-range order; with exponentially
many states this argument no longer applies, yet it is not obvious
which coupled disordered systems would order, and what such ordering
should even look like. Here we ask what happens when the local states
form a rugged, glassy landscape. We answer this in a setting where
the calculation can be carried out: copies of the Random Energy Model
(REM) \citep{derridaRandomenergyModelLimit1980,derridaRandomenergyModelExactly1981},
each with $N$ variables and the same $2^{N}$ energies, arranged
along a one--dimensional chain with coupling between variables in
nearby copies.

\begin{figure}
\begin{centering}
\includegraphics[width=1\columnwidth]{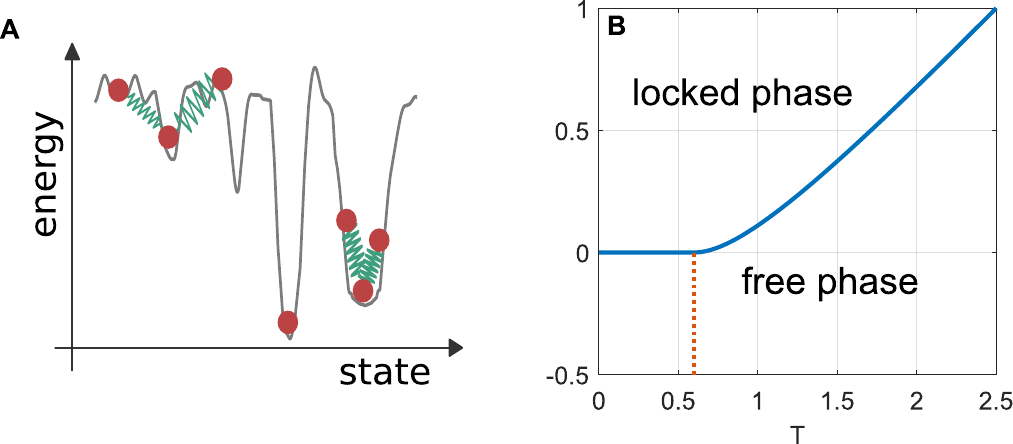}
\par\end{centering}
\caption{Disordered systems with identical disorder are coupled along a chain.
(A) Cartoon of the chain in a the energy landscape drawn in the phases,
from left to right: free phase; locked phase, with all copies at the
lowest energy state (overlapping as one bead); locking to lowest free-energy
valley. (B) Phase diagram for a long chain with state-matched coupling.
Solid line is a first-order transition between a locked phase at the
lowest energy, and a free phase, which is short-range correlated along
the chain. Dashed line is a REM freezing transition within the free
phase.}\label{fig:cartoon_phase_diagrams}
\end{figure}

Our main result is the existence of two phases: a phase with short-range
correlation along the chain, and a \emph{locked phase} where all the
copies of the system occupy the lowest energy state with probability
one, even at positive temperature, see Fig.~\ref{fig:cartoon_phase_diagrams}.
This is in contrast to the REM frozen phase, where this probability
is smaller than one. No additional phases exist, with intermediate
lengthscales between short-range and the entire system locking. The
transition between the phases is first order. Extending to the Generalized
REM (GREM) with a more structured energy landscape, we show that the
locked phase can apply either to the minimal energy state, or to a
valley with the lowest \emph{free energy}. When interactions are antiferromagnetic
along the chain, an alternating locked order can result. The same
phase diagram is obtained whether $N$ or $L$ is taken to infinity
first, though the two limits describe different situations and need
different analysis methods. For $L\gg N$, locked domains are exponentially
large in $N$, and so visible already at modest values of $N$.

In addition to the class of systems listed above \citep{narayanankuttyMetabolicCoordinationPhase2025,searInstabilitiesComplexMixtures2003,shrinivasPhaseSeparation2021,jacobsSelfAssemblyBiomolecularCondensates2021,royComplexInteractionsCan2020,pearceStabilizationExtensiveFinescale2020,buninDirectionalityCommunitylevelSelection2021,al-hiyasatSpatiotemporalNoiseStabilizes2026},
coupled copies of glass models have been used as a tool to study the
landscape of a single glass \citep{kurchanBarriersMetastableStates1993,franzRecipesMetastableStates1995,monassonStructuralGlassTransition1995,mezardHowComputeThermodynamics1999},
and in optimization algorithms on rugged landscapes \citep{baldassiUnreasonableEffectivenessLearning2016}.
In our setting the copies occupy positions in space and are coupled
only to their neighbors, so that the notion of long-range order applies.

The model is defined as follows. We consider systems formed by coupling
disordered copies \emph{with the same disorder}. We choose to work
with a one-dimensional chain, which is the most demanding test for
the existence of a phase transition. The chain length is $L$, and
the state of each copy $\mu=1,..,L$ is given by a binary vector ${\bf v}_{\mu}$
of length $N$, $v_{\mu,i}\in\left\{ +1,-1\right\} $ for $i=1..N$.

We use the REM, where the $2^{N}$ energies $E_{{\bf v}}$, are the
iid random variables, sampled from a Gaussian distribution with mean
zero and variance $NJ^{2}/2$. This is the disorder, which is the
same for all copies. The Hamiltonian reads
\begin{equation}
H_{L}=\sum_{\mu=1..L}\left[E_{{\bf v}_{\mu}}-\epsilon\,\psi\left({\bf v}_{\mu},{\bf v}_{\mu+1}\right)\right]\label{eq:H_L}
\end{equation}
where each ${\bf v}_{\mu}$ takes one of the $2^{N}$ possible binary
vectors, so that a chain state is one of $2^{NL}$. $\epsilon$ is
the coupling strength and $\psi$ is the coupling function, for which
we consider two forms. One is \emph{structured coupling}, where $\psi({\bf v},{\bf w})={\bf v}\cdot{\bf w}$.
A simpler variant is \emph{state-matched coupling}, $\psi({\bf v},{\bf w})=N\delta_{{\bf v},{\bf w}}$,
where the Kronecker delta $\delta_{{\bf v},{\bf w}}=1$ when ${\bf v}={\bf w}$,
and the prefactor $N$ is placed in the latter to make both coupling
forms extensive in $N$. Choosing the energy units to set $J=1$,
in both coupling forms, the model depends only on the parameters $T,\epsilon$.

Both couplings have instructive disorder-free limits. At $J=0$, state-matched
coupling is a Potts chain with $q=2^{N}$ states, which for $N\gg1$
has a transition at $\epsilon=T\ln2$; its ordered phase is $2^{N}$-fold
degenerate, every state being equivalent. Structured coupling at $J=0$
is $N$ independent Ising chains, with no transition.

A number of well-known results on the REM \citep{derridaRandomenergyModelExactly1981}
are used below. The number of states with energy density $u=E/N$
(in a window $\Delta u\sim N^{-\gamma}$ with $\gamma>0$) is to exponential
accuracy $e^{Ns_{1}(u)}$ with $s_{1}(u)=\ln2-u^{2}$. States exist
only where $s_{1}(u)\ge0$, that is $\left|u\right|\le u_{0}\equiv\sqrt{\ln2}$,
and $-u_{0}$ is the ground-state energy density. The free energy
of a single copy is $f_{\text{REM}}=F/N=\min_{u}\left[u-Ts_{1}(u)\right]$
subject to $s_{1}\ge0$, resulting in a second-order freezing transition,
with $f_{\text{REM}}=-T\ln2-1/\left(4T\right)$ above $T_{\text{REM}}=1/\left(2\sqrt{\ln2}\right)$,
and $f_{\text{REM}}=-u_{0}$ below it. We denote the ground state
by ${\bf v}_{\text{GS}}=\arg\min_{{\bf v}}E_{{\bf v}}$, with energy
$E_{\text{GS}}=E_{{\bf v}_{\text{GS}}}$. 

\emph{Coupling two copies}--We start with a preparatory calculation
of two copies ($L=2$) at $N\gg1$, with the state-matched coupling.
The free-energy at $(T,\epsilon)$ can be exactly calculated by separating
the partition function into cases with equal and different states
between the two copies, $Z_{2}=Z_{2}^{({\bf v}={\bf w})}+Z_{2}^{({\bf v}\ne{\bf w})}$.
Here and below, sums such as $\sum_{{\bf v}}$ run over the $2^{N}$
states of a single copy, and ${\bf v},{\bf w}$ are the states of
the two copies, so that $Z_{2}^{({\bf v}={\bf w})}=\sum_{{\bf v}}e^{-\beta\left(2E_{{\bf v}}-\epsilon N\right)}=e^{\beta\epsilon N}Z_{\text{REM}}(T/2)$,
while $Z_{2}^{({\bf v}\ne{\bf w})}=\sum_{{\bf v}\ne{\bf w}}e^{-\beta\left(E_{{\bf v}}+E_{{\bf w}}\right)}$,
giving $\left[Z_{\text{REM}}(T)\right]^{2}-Z_{\text{REM}}(T/2)$.
The free-energy density per variable, per copy, $f_{2}=\lim_{N\to\infty}\left(-\frac{T}{2N}\ln Z\right)$,
is therefore
\begin{equation}
f_{2}=\min\left[f_{\text{REM}}(T),\ f_{\text{REM}}(T/2)-\epsilon/2\right]
\end{equation}
for $\epsilon>0$, and $f_{2}=f_{\text{REM}}(T)$ for $\epsilon<0$.

\begin{figure}
\begin{centering}
\includegraphics[width=1\columnwidth]{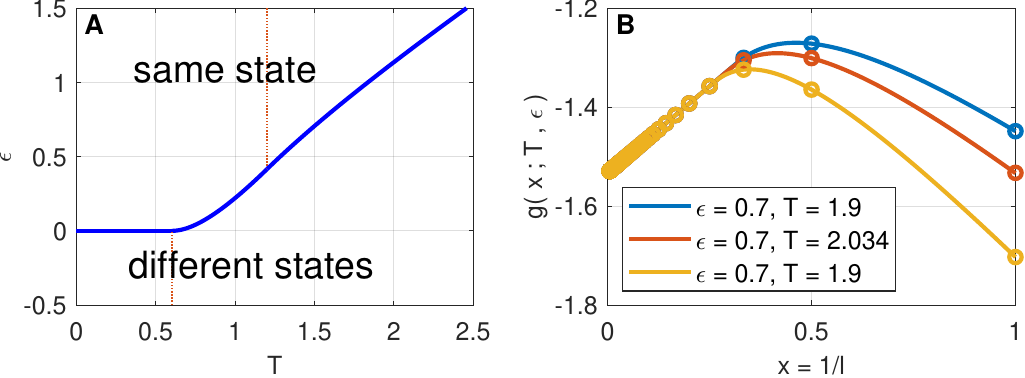}
\par\end{centering}
\caption{(A) The phase diagram of two coupled copies with state-matched coupling.
Solid line is a first order transition between same-state and different
state phases. Vertical dashed lines are REM freezing transitions within
each. (B) The free-energy density $g_{(T,\epsilon)}(x)$, with $x=1/l$.
The transition is at $g_{(T,\epsilon)}(0)=g_{(T,\epsilon)}(1)$ (here
at $T=2.034$, for $\epsilon=0.7$).}\label{fig:free-energy-density}
\end{figure}

The phase diagram divides the plane into two regions, see Fig.~\ref{fig:free-energy-density}(B).
When $f_{\text{REM}}(T)$ dominates, the system behaves like two uncoupled
copies of the REM. When $f_{\text{REM}}(T/2)-\epsilon/2$ dominates,
at large $\epsilon$ and low $T$, the two copies occupy the same
state. The factor one-half in $f_{\text{REM}}(T/2)$ enters because
each state is counted with energy $2E_{{\bf v}}$, which is equivalent
to $T\to T/2$. The transition line between the two regions is first
order (indeed, the derivatives of $f_{2}$ jump). Within each region
there is an additional freezing transition, at $T_{\text{REM}}$ and
$2T_{\text{REM}}$ for the different- and equal-state regions respectively.

\emph{Long chain limit--}Adding more copies to the chain, $L=3,4,...$,
increasingly elaborate phase diagrams can be similarly constructed.
The situation however significantly simplifies for $L\gg1$, here
taken after $N\to\infty$ of the single system size.

Divide an open chain of length $L$ into domains of length $\left\{ l_{1},l_{2},..,l_{k}\right\} $
with $L=\sum_{i=1}^{k}l_{i}$, where within each domain all copies
assume the same state, and the state is different when passing from
one domain to the next. We first treat the case $\epsilon>0$. Generalizing
the argument from the previous section, the partition function of
a domain of length $l$ is
\begin{equation}
e^{\beta\epsilon N(l-1)}\sum_{{\bf v}}e^{-l\beta Nu_{{\bf v}}}=e^{\beta\epsilon N(l-1)}Z_{\text{REM}}\left(T/l\right)\label{eq:Z_domain}
\end{equation}
Here $u_{{\bf v}}=E_{{\bf v}}/N$. The equality identifies the domain
with a single copy at the reduced temperature $T/l$. The free energy
density $f=-\frac{T}{NL}\ln Z$ constrained to the partition $\left\{ l_{i}\right\} $
reads
\begin{equation}
f_{\left\{ l_{i}\right\} }=\lim_{N\to\infty}-\frac{T}{LN}\ln\left[\prod_{i=1..k}e^{\beta N\left[\epsilon(l_{i}-1)+\zeta\right]}Z_{\text{REM}}\left(T/l_{i}\right)\right]\label{eq:f_N_gg_L}
\end{equation}
For future use, we included a domain-boundary term $\zeta$, which
for state-matched coupling discussed presently is $\zeta=0$. For
$N\to\infty$ this becomes
\begin{equation}
f_{\left\{ l_{i}\right\} }=\frac{1}{L}\sum_{i=1}^{k}l_{i}g_{(T,\epsilon)}(1/l_{i})
\end{equation}
where $g_{(T,\epsilon)}(x)$ is the per copy free-energy density along
the chain
\begin{equation}
g_{(T,\epsilon)}(x)\equiv\epsilon\left(x-1\right)-\zeta x+f_{\text{REM}}\left(xT\right)\label{eq:g_x}
\end{equation}

For any fixed $L$, one must now maximize $f_{\left\{ l_{i}\right\} }$
over all partitions $\left\{ l_{i}\right\} $ (since $N\to\infty$
was already taken). At $L\gg1$, $f$ will therefore be dominated
by sections of length $l^{*}=1/x^{*}$, with $x^{*}$ that minimizes
$g_{(T,\epsilon)}(x)$ on $0<x\le1$. $g_{(T,\epsilon)}(x)$ is always
concave, since $f_{\text{REM}}(T)$ is concave (see its expression
above), and so has a minimum at either endpoint
\begin{equation}
f=\min\left[g_{(T,\epsilon)}(0),\ g_{(T,\epsilon)}(1)\right]\label{eq:f_min_g0_g1}
\end{equation}
corresponding to $l^{*}=1$ or $l^{*}=\infty$. There are therefore
two phases, see Fig.~\ref{fig:cartoon_phase_diagrams}(B). For $l^{*}=1$
the copies are uncorrelated and $f_{L\to\infty}=f_{\text{REM}}(T)$.
For $l^{*}=\infty$ all copies assume the same state, and $f_{L}=f_{\text{REM}}(T/L)\to f_{\text{REM}}(0)$:
the chain occupies the lowest-energy state with probability one. This
is in contrast to the frozen phase of a single REM, where the ground
state carries only a finite probability $p<1$. This implies long-range
order of the individual variables $i=1..N$, with $v_{\mu,i}$ locked
to its value in the ground state.

The transition line is first-order (the derivative of $f$ jumps),
and using $g_{(T,\epsilon)}(0)=g_{(T,\epsilon)}(1)$ it is at
\begin{equation}
\epsilon_{c}(T)=T\ln2-\sqrt{\ln2}+\frac{1}{4T}\label{eq:ferro_trans_simple_coupling}
\end{equation}
for $T>T_{\text{REM}}$ and $\epsilon_{c}=0$ for $T<T_{\text{REM}}$.

Extending the above argument to $\epsilon<0$, the situation is similar
to the $L=2$ case. The free phase continues across $\epsilon=0$:
neighboring copies now occupy different states, but there are exponentially
many likely states so this is a negligible effect. Crossing to $T<T_{\text{REM}}$,
each copy chooses one of the $O(N^{0})$ low-lying states, all with
energy density $E_{\text{GS}}/N=-u_{0}$ of the REM.

\emph{Structured coupling--}We turn to the structured coupling case,
continuing with $N\gg L\gg1$. For states ${\bf v}_{\mu},{\bf v}_{\mu+1}$,
define the overlap $\alpha\equiv\left({\bf v}_{\mu}\cdot{\bf v}_{\mu+1}\right)/N$.
Given ${\bf v}_{\mu}$, the number of states ${\bf v}_{\mu+1}$ at
a prescribed overlap $\alpha$ with it, is $\binom{N}{N(1+\alpha)/2}=e^{N\left[\ln2+\phi(\alpha)\right]}$
with $\phi(\alpha)\equiv-\frac{1}{2}\left[(1-\alpha)\ln(1-\alpha)+(1+\alpha)\ln(1+\alpha)\right]$.
Therefore the density of states at given $(E_{{\bf v}_{\mu}},E_{{\bf v}_{\mu+1}},\alpha)$
is to exponential accuracy $e^{N[s_{1}(u_{\mu})+s_{1}(u_{\mu+1})+\phi(\alpha)]}$,
where $u_{\mu}=E_{{\bf v}_{\mu}}/N$.

For $L$ copies, consider equal-state domains of length $\left\{ l_{1},l_{2},..,l_{k}\right\} $
with $L=\sum_{i=1}^{k}l_{i}$ as before, and fix the overlaps $\{\alpha_{1},\alpha_{2},..\}$
between the domains. The states ${\bf v}_{i=1..k}$ for the domains
can be chosen successively from $1$ up to $k$, with each domain
after the first adding a factor of $e^{N\left[\ln2+\phi(\alpha_{i})\right]}$.
This is instead of $2^{N}$ for the state-matched case, and the $2^{N}$
is already included in $Z_{\text{REM}}$. Thus, the only changes from
the state-matched case are that $e^{\beta N\epsilon\sum_{i}(l_{i}-1)}$
is replaced by $e^{\beta N\sum_{i}\left[\epsilon(l_{i}-1)+T\phi(\alpha_{i})+\epsilon\alpha_{i}\right]}$,
with $\epsilon N\alpha_{i}$ the energetic contribution between domains.
This treats the $k$ domain energies as independent draws: indeed,
the correction from two domains that carry the same state is negligible,
as shown below. The $\alpha_{i}$ are not yet fixed, and the saddle
point picks the maximum of $\sum_{i}\left[T\phi(\alpha_{i})+\epsilon\alpha_{i}\right]$,
giving $\alpha_{i}=\alpha^{*}=\tanh(\beta\epsilon)$ and so $T\phi(\alpha^{*})+\epsilon\alpha^{*}=T\ln\left[\cosh(\beta\epsilon)\right]$.
Therefore the free energy $f$ is Eq.~(\ref{eq:f_N_gg_L}), now with
\begin{equation}
\zeta=T\ln\left[\cosh\left(\beta\epsilon\right)\right].
\end{equation}
At this saddle the entropy of available states is maximized away from
its boundaries, so no Lagrange multiplier is needed. A single domain
then has $e^{Ns_{1}(u^{*})}$ states available to it with $s_{1}(u^{*})>0$.
Therefore, repeats of states among the $k$ domains are suppressed
by $O(k^{2}e^{-Ns_{1}})$ and are negligible.

The function $g_{(T,\epsilon)}(x)$, Eq.~(\ref{eq:g_x}), is still
concave in $x$, so the phase transition line is again fixed by Eq.~(\ref{eq:f_min_g0_g1}).
This gives
\begin{equation}
\epsilon_{c}=-\frac{T}{2}\ln\left[\exp\left(\frac{\sqrt{\ln2}}{T}-\frac{1}{4T^{2}}\right)-1\right]\label{eq:structured_epsilon_c}
\end{equation}
for $T>T_{\text{REM}}$, and $\epsilon_{c}=0$ for $T<T_{\text{REM}}$,
see Fig.~\ref{fig:Phase-diagrams}(A).

The disorder therefore plays a different role in the two cases. For
state-matched coupling it breaks the $2^{N}$-fold degeneracy of the
Potts chain, selecting ${\bf v}_{{\rm GS}}$ and moving the line to
Eq.~(\ref{eq:ferro_trans_simple_coupling}), still linear at large
$T$. For structured coupling it creates the transition, and the line
rises faster than linearly, $\epsilon_{c}\simeq\frac{T}{2}\ln\frac{T}{u_{0}}$.

\begin{figure}
\begin{centering}
\includegraphics[width=1\columnwidth]{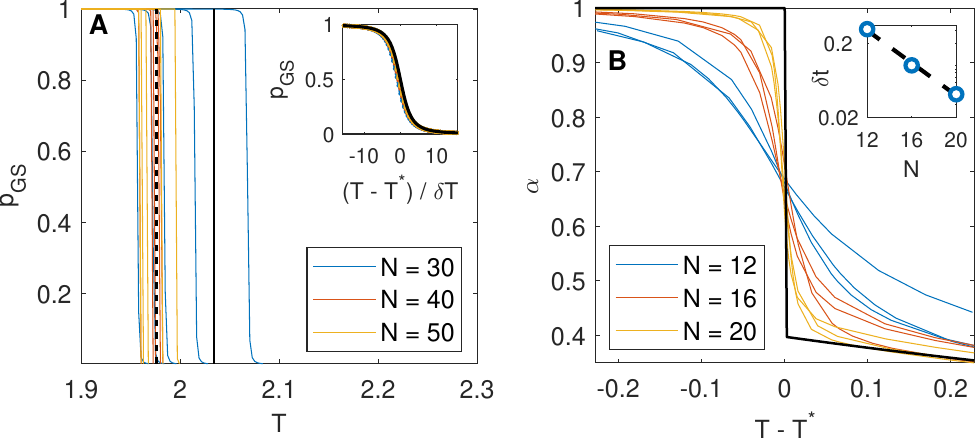}
\par\end{centering}
\caption{Transition at $L\gg N$. (A) Long chain, state-matched coupling:
the transition at $\epsilon=0.7$. $p_{\text{GS}}$ is the probability
of the minimal energy state. Every line is a single realization of
the disorder. The vertical lines are $T$ at the transition by Eq.
(\ref{eq:structured_epsilon_c}) (solid line) and corrected for $N=50$
(dashed line). Inset: after shifting by each realization's transition
temperature $T^{*}$ and rescaling by $\delta T\equiv T^{*}e^{-\epsilon N/2T^{*}}/N$,
the curves collapse, showing that they are exponentially narrow in
$N$. (B) Structured couplings: the overlap $\alpha$ between neighbors,
after shifting by $T^{*}$. Solid line: the predicted value $\alpha^{*}(T,\epsilon=0.7)$.
Inset: the transition width, measured here from the steepest slope,
$\delta T\equiv1/\left[d\alpha/dT\right]_{\text{max}}$, decreases
exponentially in $N$.}\label{fig:Long-chain-simple-coupling}
\end{figure}

\begin{figure}
\centering{}\includegraphics[width=1\columnwidth]{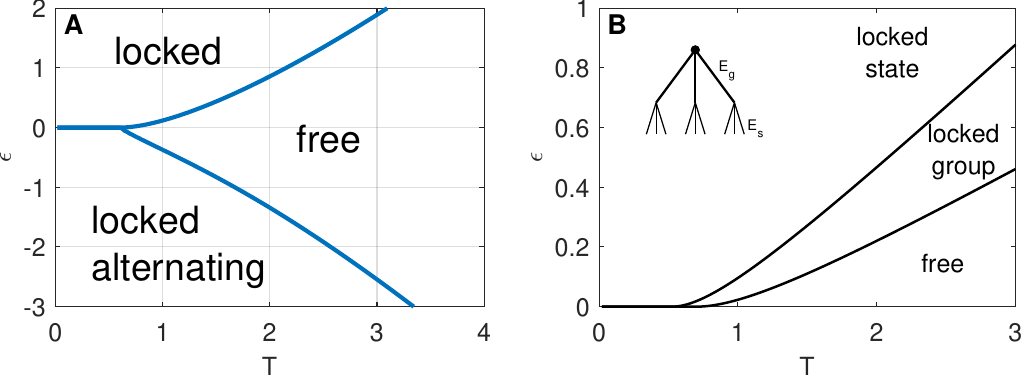}\caption{Phase diagrams. (A) REM with structured coupling. (B) GREM with $\alpha_{g}=1.3,a_{g}=0.52$.}\label{fig:Phase-diagrams}
\end{figure}

\emph{The $L\gg N\gg1$ limit--}Interestingly, the same transition
line at $\epsilon>0$ is obtained in the opposite limit, $L\gg N\gg1$,
shown analytically for state-matched coupling. Writing $Z=\text{Tr}\left(M^{L}\right)$
with $M$ the $2^{N}\times2^{N}$ transfer matrix \citep{kardarStatisticalPhysicsFields2007},
at $L\to\infty$ one has $f=-\left(T/N\right)\ln\lambda_{\max}$ with
$\lambda_{\max}$ its largest eigenvalue. For state-matched coupling
$M$ is the sum of a diagonal part $D_{{\bf vv}}=e^{-\beta E_{{\bf v}}}(e^{\beta\epsilon N}-1)$
and a rank-one part $Q_{{\bf vw}}=e^{-\beta\left(E_{{\bf v}}+E_{{\bf w}}\right)/2}$,
and $\lambda_{\max}$ undergoes an eigenvalue transition \citep{baikPhaseTransitionLargest2005,benaych-georgesEigenvaluesEigenvectorsFinite2011}
between being dominated by $D$ and by $Q$. We prove \citep{supplementalMaterial}
that for $\epsilon>0$ and $N\to\infty$, Eq.~(\ref{eq:f_min_g0_g1})
holds again, the same result as for $L\gg N$, with the transition
line of Eq.~(\ref{eq:ferro_trans_simple_coupling}) again following
as before. For structured coupling the same phenomenology is found
numerically, Fig.~\ref{fig:Long-chain-simple-coupling}(B). The top
eigenvector $\vec{y}$ similarly transitions between that of $D$
and $Q$, and the probability of states is $p_{{\bf v}}=y_{{\bf v}}^{2}$
in the transfer matrix formalism. From $M\vec{y}=\lambda_{\text{max}}\vec{y}$,
$y_{{\bf v}}\propto q_{{\bf v}}/(\lambda_{\max}-D_{\mathbf{vv}})$.
This gives $p_{{\bf v}}=\delta_{\mathbf{v},\mathbf{v}_{\text{GS}}}$
and $p_{{\bf v}}\propto e^{-\beta E_{{\bf v}}}$ inside the locked
and free phases respectively, as expected.

At finite but large $N$, the transition widens to a crossover. Here
too, we find the following phenomena numerically for both forms of
coupling, see Fig.~\ref{fig:Long-chain-simple-coupling}(A,B), and
also derive it analytically \citep{supplementalMaterial} for the
state-matched coupling from the secular equation $1=\sum_{{\bf v}}e^{-\beta E_{{\bf v}}}/\left(\lambda-D_{{\bf v}{\bf v}}\right)$
(as is standard for such problems \citep{baikPhaseTransitionLargest2005,benaych-georgesEigenvaluesEigenvectorsFinite2011,bernardMeanfieldTheoryHeterogeneous2026}).
At fixed disorder sample $\left\{ E_{{\bf v}}\right\} $, the transition
width is exponentially narrow in $N$, see Fig.~\ref{fig:Long-chain-simple-coupling}.
The sample-to-sample width of $E_{\text{GS}}$ is $O(1)$ \citep{derridaRandomenergyModelExactly1981},
so the location of the transition varies between samples much more
than the width, see Fig.~\ref{fig:Long-chain-simple-coupling}(A).
To compare realizations we define a transition temperature $T^{*}$
for each realization by $g_{(T^{*},\epsilon)}(0)=g_{(T^{*},\epsilon)}(1)$,
with $f_{{\rm REM}}(xT)$ in Eq.~(\ref{eq:g_x}) evaluated with that
realization's energies, and plot each realization against $T-T^{*}$.
Separately, the size of consecutive domains of the ground-state grows
exponentially in $N$ (for the state-matched coupling, as $e^{\beta\left(2\epsilon-\epsilon_{c}\right)N}$)
inside the locked phase, consistent with domain walls between locked
segments being exponentially expensive in $N$. In summary, even modest
values of $N$ (corresponding, recall, to the number of fields) give
a sharp transition and very large locked domains. 

For structured coupling with $\epsilon<0$ we find a transition to
a locked phase, not uniform as for $\epsilon>0$, but where a pair
of states ${\bf v},{\bf v}'$ indefinitely alternate along the chain,
$({\bf v},{\bf v}',{\bf v},{\bf v}',...)$, see Fig.~\ref{fig:cartoon_phase_diagrams}(C).
This is lowest available pair $f=(u_{1}+u_{2})/2-\epsilon\alpha$
that exists, namely that has $s_{1}(u_{1})+s_{1}(u_{2})+\phi(\alpha)\ge0$.
The transition to the free phase is first order. See calculation details
in \citep{supplementalMaterial}.

\emph{States in valleys--}The locked phase described above puts the
entire chain in the single, lowest-energy state. That is special to
the REM, where even similar states have independent energies; in other
glass models states are arranged into valleys, within which individual
variables still fluctuate \citep{mezardInformationPhysicsComputation2009}.
To see what this can change, we take a minimal variant of the Generalized
Random Energy Model (GREM) \citep{derridaGeneralizationRandomEnergy1985}:
here the $2^{N}$ states are divided into $\alpha_{g}^{N}$ groups
each containing $\alpha_{s}^{N}$ states, with $\alpha_{g}\alpha_{s}=2$.
The energy of state is $E_{{\bf v}}=E_{g,g({\bf v})}+E_{s,{\bf v}}$,
where $E_{g}$ is shared by all states of group $g$, and $E_{s,{\bf v}}$
independent within it, with variances $NJ^{2}a_{g}/2$ and $N(1-a_{g})/2$,
so that $\left\langle E_{{\bf v}}^{2}\right\rangle =N/2$ as for the
REM. Generalizing the state-matched coupling,
\begin{equation}
H_{2}=E_{{\bf v}}+E_{{\bf w}}-\epsilon N\left[\delta_{g({\bf v}),g({\bf w})}+\delta_{{\bf v},{\bf w}}\right]
\end{equation}
Three phases result. Two are as for the REM: different groups (and
therefore also different states) in different copies, or the whole
chain in the lowest-energy state. In the third phase, the groups are
locked while the states inside them differ from copy to copy, so the
chain is not in any single state, and the variables thermally fluctuate
between copies. The locked group is the one of lowest \emph{free energy}:
for $l$ consecutive copies sharing a group but free to differ inside
it, summing over states gives $e^{\beta\epsilon N(l-1)}\sum_{g}e^{-l\beta\mathcal{F}_{g}}$
with $\mathcal{F}_{g}\equiv-T\ln\sum_{{\bf v}\in g}e^{-\beta E_{{\bf v}}}$,
which is the domain sum in Eq.~(\ref{eq:Z_domain}), with state energies
replaced by group free energies, so the lowest $\mathcal{F}_{g}$
is selected when $l$ is large. With equal group sizes this is the
group of lowest $E_{g}$; the entropic part of $\mathcal{F}_{g}$
enters when groups differ in size. We prove that, for $a_{g}/\ln\alpha_{g}>a_{s}/\ln\alpha_{s}$,
these are the only phases, with intermediate domain lengths excluded
\citep{supplementalMaterial}. The intermediate phase appears whenever
$\alpha_{g}<\alpha_{s}$, that is, fewer groups than states within
them.

\emph{Discussion--}Here we studied coupled glassy models, as a model
of disordered interactions of many fields. We obtained a locked phase,
producing a unique form of long-range order. Extending these results,
it would be interesting to study the model in higher dimensions, in
particular due to the possible appearance of ferromagnetic order.
Another important question is what happens in glass models other than
the REM. Coupled copies of glass models, analyzed to study properties
of the energy landscape \citep{kurchanBarriersMetastableStates1993,franzRecipesMetastableStates1995,monassonStructuralGlassTransition1995,mezardHowComputeThermodynamics1999},
serve as a promising starting point. The REM itself is a limit of
a family of glass models \citep{derridaRandomenergyModelExactly1981},
establishing the results here as a limiting case. The results point
to the importance of landscape structure, as the overlaps between
valleys and the heights of barriers that separate them \citep{kurchanBarriersMetastableStates1993,mezardHowComputeThermodynamics1999,franzRecipesMetastableStates1995}
affect the domain walls. Valleys are organized hierarchically in some
cases \citep{mezardInformationPhysicsComputation2009} as stylized
in the GREM \citep{derridaGeneralizationRandomEnergy1985}.

A quantum Hamiltonian maps to a classical chain by the Suzuki-Trotter
formula, and so the REM in a transverse field \citep{goldschmidtSolvableModel1990}
maps to a chain like Eq.~(\ref{eq:H_L}). However, the many-layer
limit in the quantum prescription corresponds to a joint limit of
large $(\epsilon,T,L)$ which seems unnatural in the present setting,
and leads to a different phenomenology, without a locked phase at
$T>0$ \citep{supplementalMaterial}. It would be interesting to see
if other relations to quantum systems would be fruitful.

Finally, the dynamics of such systems, involving an interplay of glassy
dynamics and coarsening \citep{buninEvolutionaryFeatures2025} are
a fascinating subject for future research.

\selectlanguage{english}%
\bibliographystyle{unsrt}
\bibliography{my_library,manual_refs}

\begin{thebibliography}{10}

\bibitem{narayanankuttyMetabolicCoordinationPhase2025}
Krishnadev Narayanankutty, Jos{\'e}~Antonio {Pereiro-Morejon}, Ari{\'a}n
  {Ferrero-Fern{\'a}ndez}, Valentina Onesto, Stefania Forciniti, Loretta~L.
  Del~Mercato, Roberto Mulet, Andrea De~Martino, David~S. Tourigny, and Daniele
  De~Martino.
\newblock Metabolic coordination and phase transitions in spatially distributed
  multi-cellular systems.
\newblock {\em Communications Physics}, 8(1):205, May 2025.

\bibitem{searInstabilitiesComplexMixtures2003}
Richard~P. Sear and Jos{\'e}~A. Cuesta.
\newblock Instabilities in {{Complex Mixtures}} with a {{Large Number}} of
  {{Components}}.
\newblock {\em Physical Review Letters}, 91(24):245701, December 2003.

\bibitem{shrinivasPhaseSeparation2021}
Krishna Shrinivas and Michael~P. Brenner.
\newblock Phase separation in fluids with many interacting components.
\newblock {\em Proceedings of the National Academy of Sciences},
  118(45):e2108551118, November 2021.

\bibitem{jacobsSelfAssemblyBiomolecularCondensates2021}
William~M. Jacobs.
\newblock Self-{{Assembly}} of {{Biomolecular Condensates}} with {{Shared
  Components}}.
\newblock {\em Physical Review Letters}, 126(25):258101, June 2021.

\bibitem{royComplexInteractionsCan2020}
Felix Roy, Matthieu Barbier, Giulio Biroli, and Guy Bunin.
\newblock Complex interactions can create persistent fluctuations in
  high-diversity ecosystems.
\newblock {\em PLOS Computational Biology}, 16(5):e1007827, May 2020.

\bibitem{pearceStabilizationExtensiveFinescale2020}
Michael~T. Pearce, Atish Agarwala, and Daniel~S. Fisher.
\newblock Stabilization of extensive fine-scale diversity by ecologically
  driven spatiotemporal chaos.
\newblock {\em Proceedings of the National Academy of Sciences},
  117(25):14572--14583, June 2020.

\bibitem{buninDirectionalityCommunitylevelSelection2021}
Guy Bunin.
\newblock Directionality and community-level selection.
\newblock {\em Oikos}, 130(4):489--500, April 2021.

\bibitem{al-hiyasatSpatiotemporalNoiseStabilizes2026}
Amer {Al-Hiyasat}, Daniel~W. Swartz, Jeff Gore, and Mehran Kardar.
\newblock Spatiotemporal noise stabilizes unbounded diversity in
  strongly-competitive communities, August 2026.

\bibitem{wignerRandomMatricesPhysics1967}
Eugene~P. Wigner.
\newblock Random {{Matrices}} in {{Physics}}.
\newblock {\em SIAM Review}, 9(1):1--23, January 1967.

\bibitem{mayWillLargeComplex1972}
Robert~M. May.
\newblock Will a large complex system be stable?
\newblock {\em Nature}, 238(5364):413--414, 1972.

\bibitem{derridaRandomenergyModelLimit1980}
Bernard Derrida.
\newblock Random-energy model: {{Limit}} of a family of disordered models.
\newblock {\em Physical Review Letters}, 45(2):79, 1980.

\bibitem{derridaRandomenergyModelExactly1981}
Bernard Derrida.
\newblock Random-energy model: {{An}} exactly solvable model of disordered
  systems.
\newblock {\em Physical Review B}, 24(5):2613, 1981.

\bibitem{kurchanBarriersMetastableStates1993}
J.~Kurchan, G.~Parisi, and M.~A. Virasoro.
\newblock Barriers and metastable states as saddle points in the replica
  approach.
\newblock {\em Journal de Physique I}, 3(8):1819--1838, August 1993.

\bibitem{franzRecipesMetastableStates1995}
Silvio Franz and Giorgio Parisi.
\newblock Recipes for {{Metastable States}} in {{Spin Glasses}}.
\newblock {\em Journal de Physique I}, 5(11):1401--1415, November 1995.

\bibitem{monassonStructuralGlassTransition1995}
R{\'e}mi Monasson.
\newblock Structural {{Glass Transition}} and the {{Entropy}} of the
  {{Metastable States}}.
\newblock {\em Physical Review Letters}, 75(15):2847--2850, October 1995.

\bibitem{mezardHowComputeThermodynamics1999}
Marc M{\'e}zard.
\newblock How to compute the thermodynamics of a glass using a cloned liquid.
\newblock {\em Physica A: Statistical Mechanics and its Applications},
  265(3-4):352--369, April 1999.

\bibitem{baldassiUnreasonableEffectivenessLearning2016}
Carlo Baldassi, Christian Borgs, Jennifer~T. Chayes, Alessandro Ingrosso, Carlo
  Lucibello, Luca Saglietti, and Riccardo Zecchina.
\newblock Unreasonable effectiveness of learning neural networks: {{From}}
  accessible states and robust ensembles to basic algorithmic schemes.
\newblock {\em Proceedings of the National Academy of Sciences},
  113(48):E7655--E7662, November 2016.

\bibitem{kardarStatisticalPhysicsFields2007}
Mehran Kardar.
\newblock {\em Statistical {{Physics}} of {{Fields}}}.
\newblock Cambridge University Press, 1 edition, June 2007.

\bibitem{baikPhaseTransitionLargest2005}
Jinho Baik, G{\'e}rard Ben~Arous, and Sandrine P{\'e}ch{\'e}.
\newblock Phase transition of the largest eigenvalue for nonnull complex sample
  covariance matrices.
\newblock {\em The Annals of Probability}, 33(5), September 2005.

\bibitem{benaych-georgesEigenvaluesEigenvectorsFinite2011}
Florent {Benaych-Georges} and Raj~Rao Nadakuditi.
\newblock The eigenvalues and eigenvectors of finite, low rank perturbations of
  large random matrices.
\newblock {\em Advances in Mathematics}, 227(1):494--521, May 2011.

\bibitem{supplementalMaterial}
See Supplemental Material for the transfer-matrix treatment of the $L\gg N$
  limit, the finite-$N$ crossover, and the numerical methods.

\bibitem{bernardMeanfieldTheoryHeterogeneous2026}
Maximilien Bernard, Jean-Philippe Bouchaud, and Pierre Le~Doussal.
\newblock Mean-field theory for heterogeneous random growth with
  redistribution.
\newblock {\em Physical Review E}, 113(3):L032101, March 2026.

\bibitem{mezardInformationPhysicsComputation2009}
Marc M{\'e}zard and Andrea Montanari.
\newblock {\em Information, {{Physics}}, and {{Computation}}}.
\newblock OUP Oxford, January 2009.

\bibitem{derridaGeneralizationRandomEnergy1985}
B.~Derrida.
\newblock A generalization of the {{Random Energy Model}} which includes
  correlations between energies.
\newblock {\em Journal de Physique Lettres}, 46(9):401--407, 1985.

\bibitem{goldschmidtSolvableModel1990}
Yadin~Y. Goldschmidt.
\newblock Solvable model of the quantum spin glass in a transverse field.
\newblock {\em Physical Review B}, 41(7):4858--4861, March 1990.

\bibitem{buninEvolutionaryFeatures2025}
Guy Bunin and Olivier Rivoire.
\newblock Evolutionary features in a minimal physical system: {{Diversity}},
  selection, growth, inheritance, and adaptation.
\newblock {\em Proceedings of the National Academy of Sciences},
  122(31):e2425753122, August 2025.

\bibitem{bovierFluctuationsFreeEnergy2002}
Anton Bovier, Irina Kurkova, and Matthias L{\"o}we.
\newblock Fluctuations of the free energy in the {{REM}} and the \$p\$-spin
  {{SK}} models.
\newblock {\em The Annals of Probability}, 30(2), April 2002.

\end{thebibliography}

\newpage{}

\selectlanguage{american}%
\onecolumngrid
\begin{center}
\textbf{Supplemental Material for ``Locking transition in coupled
disordered systems''}
\par\end{center}

\section{The long chain, $L\gg N$}

Here we treat the chain in the $L\to\infty$ at fixed $N$, and then
$N\to\infty$. The free energy follows from a transfer matrix.

\subsection{Transfer matrix formulation}

\begin{figure}[b]
\begin{centering}
\includegraphics[width=0.5\textwidth]{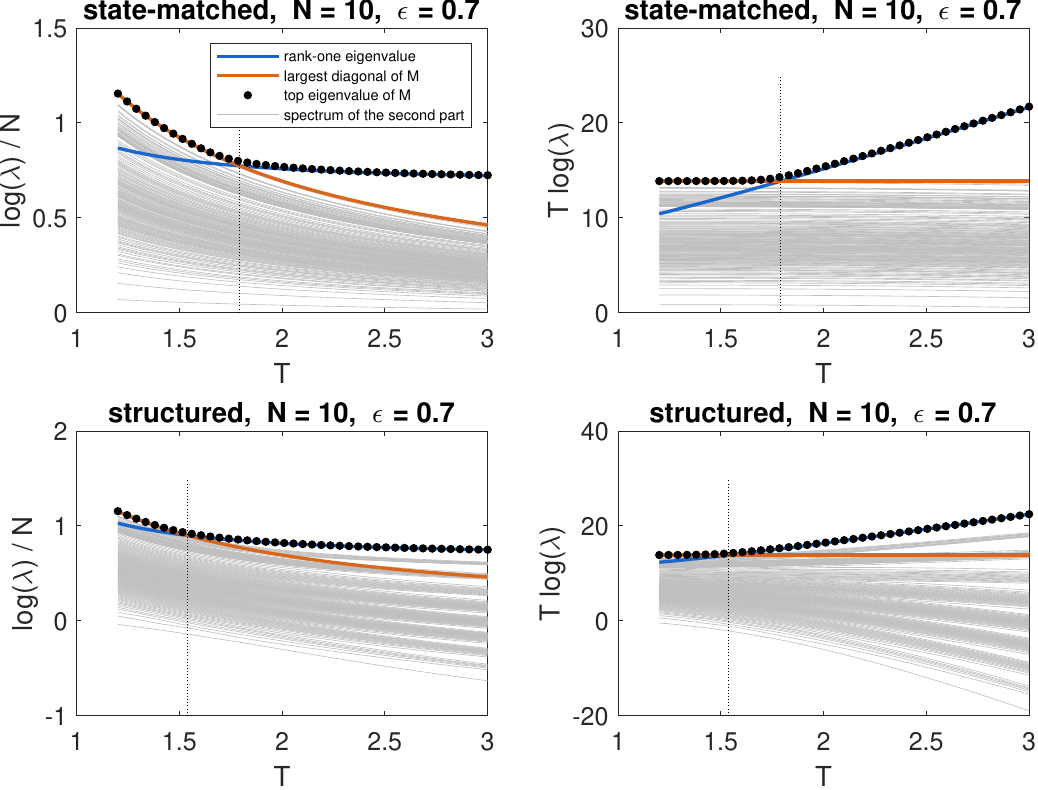}
\par\end{centering}
\caption{The eigenvalue transition. Top: state matched coupling, shown as
a function of $N^{-1}\ln\lambda$ (left) and $T\ln\lambda$ (right).
Bottom row: same for structured coupling. The thinner lines show the
spectra of $D$ for state-matched coupling, and $B$ for structured
coupling. Vertical line shows the theoretical position of the transition.}\label{fig:The-eigenvalue-transition.}
\end{figure}

For the state-matched coupling, $\psi({\bf v},{\bf w})=N\delta_{{\bf v},{\bf w}}$
the partition function of a periodic chain of $L$ copies is $Z_{L}=\text{Tr}\,M^{L}$
with the $2^{N}\times2^{N}$ transfer matrix 
\begin{equation}
M_{{\bf v}{\bf w}}=\exp\left[-\beta\left(E_{{\bf v}}+E_{{\bf w}}\right)/2+\beta\epsilon N\delta_{{\bf v},{\bf w}}\right]\label{eq:sm_M}
\end{equation}
so that $f=-\frac{T}{N}\ln\lambda_{1}$ as $L\to\infty$ at fixed
$N$, where $\lambda_{1}$ is the largest eigenvalue. Writing $q_{{\bf v}}\equiv e^{-\beta E_{{\bf v}}/2}$,
the matrix splits exactly into a rank-one part and a diagonal part,
\begin{align}
M & =Q+D\\
Q_{{\bf v}{\bf w}} & =q_{{\bf v}}q_{{\bf w}}\\
D_{{\bf v}{\bf w}} & =\left(e^{\beta\epsilon N}-1\right)q_{{\bf v}}^{2}\,\delta_{{\bf v},{\bf w}}\label{eq:sm_split}
\end{align}
The eigenvalues of $Q$ are $\sum_{{\bf v}}q_{{\bf v}}^{2}=Z_{\text{REM}}(T)$
and $0$ with multiplicity $2^{N}-1$; those of $D$ are its diagonal
entries. A rank-one perturbation of a matrix with a given spectrum
is the setting of the Baik-Ben Arous-Peché (BBP) transition \citep{baikPhaseTransitionLargest2005,benaych-georgesEigenvaluesEigenvectorsFinite2011}
and similar eigenvalue transitions, and the two phases below are its
two sides. The derivation below is, however, self-contained. Fig.
\ref{fig:The-eigenvalue-transition.} shows the transition in one
realization, for both types of coupling.

\subsection{Calculating observables from the transfer matrix}

The eigenvector gives the statistics of the chain directly. The probability
that the copy at position $k$ occupies state ${\bf v}$ is 
\begin{equation}
P({\bf v})=\frac{1}{Z_{L}}\text{Tr}\left[\delta_{\sigma_{k},{\bf v}}M^{L}\right]=x_{{\bf v}}^{2}\label{eq:sm_Pv}
\end{equation}
in the limit $L\to\infty$, normalized to one because ${\bf x}$ is.
The joint probability of two neighboring copies is 
\begin{equation}
P\left(\sigma_{k}={\bf v},\sigma_{k+1}={\bf w}\right)=M_{{\bf v}{\bf w}}\,x_{{\bf v}}x_{{\bf w}}/\lambda_{1}\label{eq:sm_Pvw}
\end{equation}
from which the probability that the chain stays in the ground state
from one copy to the next is 
\begin{equation}
\text{Prob}\left(\sigma_{k+1}={\bf v}_{\text{GS}}\mid\sigma_{k}={\bf v}_{\text{GS}}\right)=M_{{\bf v}_{\text{GS}}{\bf v}_{\text{GS}}}/\lambda_{1}\label{eq:sm_cond}
\end{equation}
The quantity plotted in the main text is $p_{{\bf v}}$ of Eq. (\ref{eq:sm_Pv});
the conditional probability above instead sets the ordering length
along the chain, since staying locked over $k$ steps has probability
$\left[M_{{\bf v}_{\text{GS}}{\bf v}_{\text{GS}}}/\lambda_{1}\right]^{k}$.

\subsection{The maximal eigenvalue of $M=D+Q$}

The result for the maximal eigenvalues of $M=D+Q$ can also be obtained
from the secular equation discussed in the next subsection. Here it
is derived from a shorter and more direct argument.

Since $Q$ is rank one and $D$ is diagonal
\[
\lambda_{\max}^{(D)}=e^{\beta|E_{{\rm GS}}|}(e^{\beta\epsilon N}-1)\ \ \ ;\ \ \ \lambda_{\max}^{(Q)}=\sum_{{\bf v}}q_{{\bf v}}^{2}=Z_{{\rm REM}}(T)
\]
and with $\Lambda^{(D,Q)}\equiv\lim_{N\to\infty}\left[N^{-1}\ln\lambda_{\max}^{(D,Q)}\right]$,
\[
\Lambda^{(D)}=\beta\left(u_{0}+\epsilon\right)\ \ \ ;\ \ \ \Lambda^{(Q)}=-\beta f_{\text{REM}}(T)
\]
Both $D$ and $Q$ are positive semidefinite for $\epsilon>0$, so
\[
\max\{\lambda_{\max}^{(Q)},\lambda_{\max}^{(D)}\}\le\lambda_{\max}\le\lambda_{\max}^{(Q)}+\lambda_{\max}^{(D)}\ .
\]
Both bounds follow from $\lambda_{\max}=\max_{|x|=1}x^{\top}Mx$.
The bounds on $\lambda_{\max}$ differ by at most a factor of two,
therefore $f=\lim_{N\to\infty}-\left(T/N\right)\ln\lambda_{\max}=-T\max\left[\Lambda^{(D)},\ \Lambda^{(Q)}\right]=\min\left[g_{(T,\epsilon)}(0),g_{(T,\epsilon)}(1)\right]$
given in the main text.

\subsection{The largest eigenvalue, using the secular equation}

An eigenvector of $M=Q+D$ obeys $\left(\lambda-D\right){\bf x}={\bf q}\left({\bf q}\cdot{\bf x}\right)$.
Provided ${\bf q}\cdot{\bf x}\neq0$, taking the inner product with
${\bf q}$ gives the secular equation 
\begin{equation}
G(\lambda)\equiv\sum_{{\bf v}}\frac{q_{{\bf v}}^{2}}{\lambda-D_{{\bf v}{\bf v}}}=1,\qquad x_{{\bf v}}\propto\frac{q_{{\bf v}}}{\lambda-D_{{\bf v}{\bf v}}}\label{eq:sm_secular}
\end{equation}
The function $G$ decreases monotonically from $+\infty$ to $0$
on $\left(\max_{{\bf v}}D_{{\bf v}{\bf v}},\infty\right)$, so it
has exactly one value where $G(\lambda)=1$ there. That solution is
the largest eigenvalue. By the second relation in Eq. (\ref{eq:sm_secular})
its eigenvector has all entries of one sign, whereas a solution lying
between two poles gives components of both signs; by the Perron-Frobenius
theorem the eigenvector of the largest eigenvalue of a positive matrix
is the only one without sign changes. The probability that a copy
occupies state ${\bf v}$ is then $p_{{\bf v}}=x_{{\bf v}}^{2}$ with
${\bf x}$ normalized.

\subsection{Sharpness of the transition at finite $N$}

At finite $N$ the two branches are joined smoothly, and Eq. (\ref{eq:sm_secular})
gives the crossover in closed form. The transition occurs when the
top eigenvalue approaches the ground-state pole, so set $\delta\equiv\lambda_{\max}-D_{0}$
with $D_{0}\equiv\max_{{\bf v}}D_{{\bf v}{\bf v}}$. Separating the
ground state from the sum, 
\begin{equation}
1=\frac{e^{-\beta E_{\text{GS}}}}{\delta}+S_{>0}\label{eq:sm_split_gs}
\end{equation}
To evaluate $S_{>0}$ split it at $k=N^{\alpha}$. The first $k$
states have gaps of order $O(N^{0})$ and contribute at most $O(N^{\alpha}e^{-\beta N\epsilon})$,
negligible. For the bulk, $D_{{\bf v}{\bf v}}\ll D_{0}$, so the denominator
may be taken as $\lambda_{\max}-D_{{\bf v}{\bf v}}\simeq\lambda_{\max}=D_{0}+\delta$,
giving 
\begin{equation}
1=\frac{e^{-\beta E_{\text{GS}}}}{\delta}+\frac{Z_{\text{REM}}(T)}{D_{0}+\delta}\label{eq:sm_uniform}
\end{equation}
Two parameters are relevant,
\begin{equation}
w^{2}\equiv\frac{e^{-\beta E_{\text{GS}}}}{D_{0}}=e^{-\beta\epsilon N},\qquad\eta\equiv\frac{Z_{\text{REM}}(T)}{D_{0}}\label{eq:sm_eta_w}
\end{equation}
The ratio $\eta$ is the control parameter: $N^{-1}\ln\eta=-\beta f_{\text{REM}}(T)-\beta\left(u_{0}+\epsilon\right)$
that vanishes on the transition line, Eq.~(8) of the main text, so
$\eta>1$ in the free phase and $\eta<1$ in the locked phase. $w$
is exponentially small in $N$, and will set the width of the transition
in $\eta$.

Define the rescaled gap $y\equiv\delta/D_{0}$. Multiplying Eq. (\ref{eq:sm_uniform})
by $\delta\left(D_{0}+\delta\right)$ and dropping $w^{2}y$ against
$y$, 
\begin{equation}
y^{2}+\left(1-\eta\right)y-w^{2}=0\label{eq:sm_quad_eta}
\end{equation}
whose positive root, the one with $\delta>0$, is 
\begin{equation}
y\left(\eta\right)=\frac{\left(\eta-1\right)+\sqrt{\left(\eta-1\right)^{2}+4w^{2}}}{2}\label{eq:sm_y_eta}
\end{equation}
Away from the transition $\left|\eta-1\right|\gg w$ and $y\simeq\eta-1$
in the free phase, where the eigenvalue sits well above $D_{0}$,
and $y\simeq w^{2}/\left(1-\eta\right)$ in the locked phase, where
it is pinned to it. The crossover between them happens in $\left|\eta-1\right|\lesssim2w$.
As inside the window, $\eta-1\simeq\ln\eta$ and $\partial\left(N^{-1}\ln\eta\right)/\partial\epsilon=-\beta$,
this is a window in $\epsilon$ of width
\begin{equation}
\delta\epsilon\simeq\frac{2w}{\beta N}=\frac{2T}{N}e^{-\beta\epsilon N/2}\label{eq:sm_width_eta}
\end{equation}
exponentially narrow in $N$. The ground-state occupation follows
from the eigenvector, $y_{{\bf v}}\propto q_{{\bf v}}/\left(\lambda_{1}-D_{{\bf v}{\bf v}}\right)$,
as 
\begin{equation}
p_{\text{GS}}=\left[1+\frac{\eta}{w^{2}}\frac{y^{2}}{\left(1+y\right)^{2}}\right]^{-1}\label{eq:sm_pgs_eta}
\end{equation}

It remains to express this in the variable plotted in the inset of
Fig.~3(A). Near the transition $N^{-1}\ln\eta=-\beta\left[f_{\text{REM}}(T)+u_{0}+\epsilon\right]$
is linear in $T$: the bracket vanishes at $T^{*}$, where $f_{\text{REM}}=-\left(u_{0}+\epsilon\right)$,
so the derivative is $-f_{\text{REM}}^{\prime}(T^{*})/T^{*}=s_{\text{REM}}/T^{*}$,
the REM entropy per variable at the transition, and 
\begin{equation}
\eta-1\simeq\frac{N\,s_{\text{REM}}}{T^{*}}\left(T-T^{*}\right)=s_{\text{REM}}\,w\,\frac{T-T^{*}}{\delta T},\qquad\delta T\equiv\frac{T^{*}}{N}e^{-\beta\epsilon N/2}\label{eq:sm_eta_lin-1}
\end{equation}
The crossover window $\left|\eta-1\right|\lesssim2w$ is thus a window
in temperature of width $\delta T$, exponentially narrow in $N$.
Substituting into the root Eq.~(\ref{eq:sm_y_eta}) gives $y=w\,e^{\text{arcsinh}\left[\left(s_{\text{REM}}/2\right)\left(T-T^{*}\right)/\delta T\right]}$,
and with $\eta\simeq1$ and $y\ll1$ Eq.~(\ref{eq:sm_pgs_eta}) becomes
\begin{equation}
p_{\text{GS}}=\left[1+\exp\left(2\,\text{arcsinh}\left[\frac{s_{\text{REM}}}{2}\frac{T-T^{*}}{\delta T}\right]\right)\right]^{-1}\label{eq:sm_pgs_x-1}
\end{equation}
This is the solid curve in the inset of Fig.~3(A), plotted against
$\left(T-T^{*}\right)/\delta T$ with each realization contributing
its own $T^{*}$. At $\epsilon=0.7$ the transition is at $T^{*}=2.034$
and $s_{\text{REM}}/2=0.316$.

\subsubsection{Length of locked domains}

The mean length of a locked domain can be written in the variables
of Eq. (\ref{eq:sm_eta_w}). With $M_{{\bf v}_{\text{GS}}{\bf v}_{\text{GS}}}=D_{0}/\left(1-w^{2}\right)$
and $\lambda_{1}=D_{0}\left(1+y\right)$, 
\begin{equation}
L^{-1}=1-\frac{M_{{\bf v}_{\text{GS}}{\bf v}_{\text{GS}}}}{\lambda_{1}}\simeq y-w^{2}\label{eq:sm_L}
\end{equation}
Inside the locked phase $\eta<1$ and $y\simeq w^{2}/\left(1-\eta\right)$,
so that 
\begin{equation}
L\simeq\frac{1-\eta}{\eta\,w^{2}}\simeq\frac{1}{\eta\,w^{2}}\label{eq:sm_L2}
\end{equation}
Using $N^{-1}\ln\eta=-\beta f_{\text{REM}}(T)-\beta\left(u_{0}+\epsilon\right)$and
$N^{-1}\ln w^{2}=-\beta\epsilon$, 
\begin{equation}
\frac{1}{N}\ln L\simeq\beta\left(2\epsilon-\epsilon_{c}\right)\label{eq:sm_Lrate}
\end{equation}
Note that since this is a first-order phase-transition, the same qualitative
growth extends all the way to the transition line.

\begin{figure}
\centering{}\includegraphics[width=0.5\textwidth]{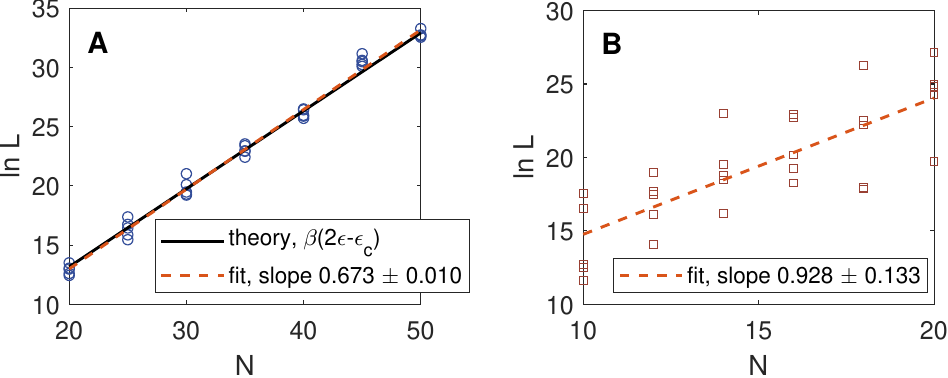}\caption{Length of domain lengths in the ground state. (A) For state-matched
coupling, at $T=1.5,\epsilon=0.7$. Linear fit of $\ln L$ versus
$N$. Theory line is Eq.~(\ref{eq:sm_Lrate}). (B) For structured
coupling, at $T=0.46,\epsilon=0.7$. Linear fit of $\ln L$ versus
$N$.}\label{fig:Length-of-domain}
\end{figure}

Fig. \ref{fig:Length-of-domain} shows the numerical results using
the transfer matrix Eq. (\ref{eq:sm_L}), for both forms of coupling.
In both coupling forms, domains display exponential growth.

\subsection{Comparing with finite-$N$ numerics}

Two finite-$N$ effects must be accounted for before the numerics
are compared with the asymptotic transition line. First, the transition
depends on the ground-state energy density, which at finite $N$ is
shallower than $u_{0}=J\sqrt{\ln2}$; using the known correction \citep{derridaRandomenergyModelExactly1981}
\begin{align}
-\frac{E_{\text{GS}}}{N}= & J\sqrt{\ln2}-\frac{T_{\text{REM}}\ln N}{2N}-\frac{T_{\text{REM}}}{2N}\ln\left(4\pi\ln2\right)-\frac{T_{\text{REM}}}{N}\Gamma^{\prime}(1)\label{eq:sm_EGS}
\end{align}
at $\epsilon=0.7$ and $J=1$ this gives transition temperatures $1.98$
at $N=50$, against the asymptotic $2.03$. Both are plotted as vertical
lines in Fig. 3(A)

Second, realizations differ. The partition function differs between
realizations by $O(N^{0})$, much more than the exponentially small
width of the transition. To compare them each realization is assigned
its own transition temperature $T^{*}$, defined by $g_{(T^{*},\epsilon)}(0)=g_{(T^{*},\epsilon)}(1)$
with $f_{\text{REM}}$ evaluated from that realization, 
\begin{equation}
\epsilon-\zeta=f_{\text{REM}}(0)-f_{\text{REM}}(T^{*})=\frac{T^{*}}{N}\ln\sum_{{\bf v}}e^{-\Delta_{{\bf v}}/T^{*}}\label{eq:sm_Tstar}
\end{equation}
Only two numbers enter, $E_{\text{GS}}$ and $\ln Z$. The scatter
in $T^{*}$ indeed comes almost entirely from the first: $\ln Z$
is self-averaging for all $T>T_{\text{REM}}$, with corrections exponentially
small in $N$ \citep{bovierFluctuationsFreeEnergy2002}, whereas $E_{\text{GS}}$
fluctuates at $O(N^{0})$. Shifting each curve by its own $T^{*}$
and rescaling by Eq. (\ref{eq:sm_width_eta}) collapses the single-realization
curves onto Eq. (\ref{eq:sm_pgs_eta}), shown in the inset of Fig.
3(A).

\subsection{Structured coupling}

\begin{figure}
\begin{centering}
\includegraphics[width=0.3\textwidth]{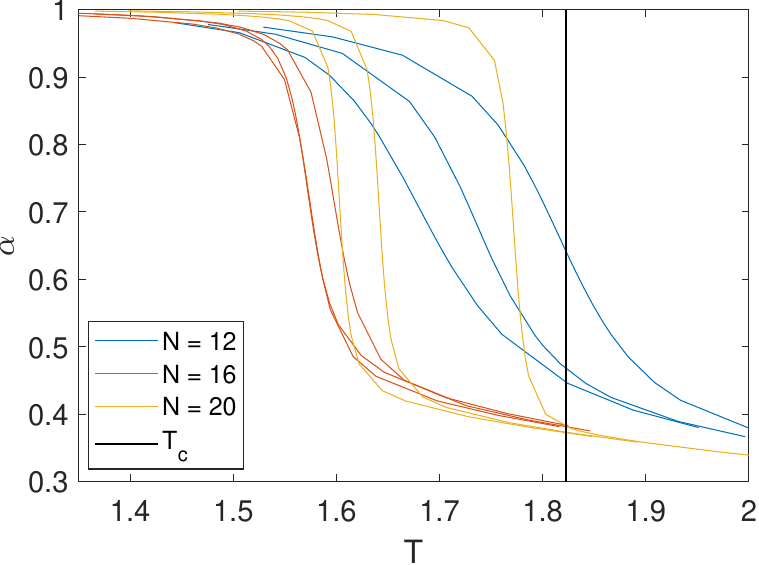}\ \ \ \ \ \ \ \ \includegraphics[width=0.3\columnwidth]{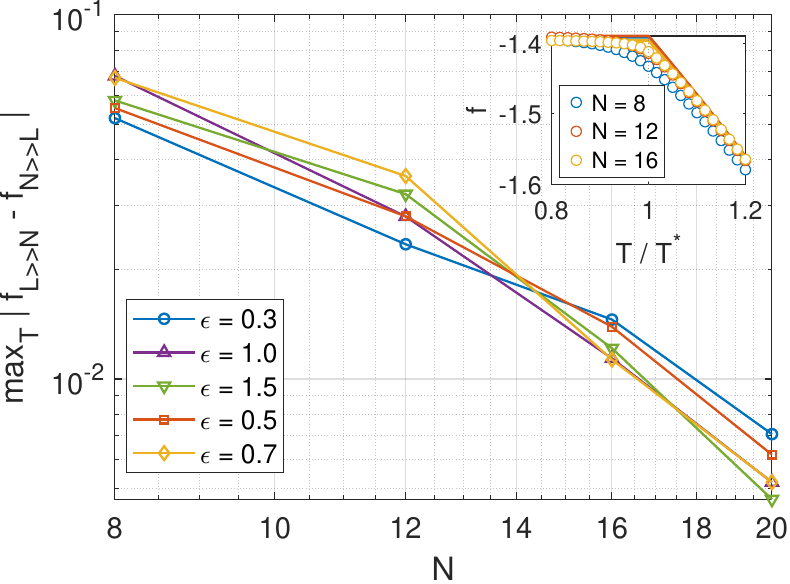}
\par\end{centering}
\caption{Left panel: Transition lines for $L\gg N$ with structured coupling.
This is the same data as Fig. 3(B) in the main text, without shifting
by the transition temperature $T^{*}$. The line is the theoretical
asymptotic result at $N\to\infty$, Eq.~(9) in the main text. Right
panel: The free energy around the transition for structured couplings
and $L\gg N$, tested against $N\gg L$ theory, Eq.~(7,9) in the
main text. The maximal difference between the two is shown as a function
of $N$ in a region around the transition. Inset: an example of the
theory curve and numerics for $\epsilon=0.7$. The position of the
transition and $f$ were chosen in a realization-dependent way, as
in Fig. (3)A in the main text.}\label{fig:Transition-lines-without-shifting}
\end{figure}

For the structured coupling, $\psi({\bf v},{\bf w})={\bf v}\cdot{\bf w}$,
the transfer matrix is no longer a rank-one perturbation of a diagonal
matrix. Numerically the transfer matrix can still be diagonalized
for $N\lesssim25$, and all the qualitative features found above for
the state-matched coupling are recovered: the transition is exponentially
narrow in $N$, its location varies between realizations and is captured
by $T^{*}$ of (\ref{eq:sm_Tstar}), and the ordering length along
the chain grows exponentially with $N$. The overlap between neighboring
copies, shifted by $T^{*}$, is shown in Fig. 3(B). The same data
is shown in Fig. \ref{fig:Transition-lines-without-shifting}(left
panel), without the shift (so, the equivalent of Fig. 3(A) but for
structured coupling).

As further tests that the transition line for structured couplings
at $L\gg N$ is the same as with $N\gg L$, we also compared the free
energy itself against theory, for a range of values of $\epsilon$,
see Fig. \ref{fig:Transition-lines-without-shifting}(right panel).

\subsubsection{A possible theoretical direction}

More theoretical insight for structured couplings can possibly be
gained by expanding the Boltzmann term of the coupling, in the parity
functions $\chi_{S}({\bf v})=\prod_{i\in S}v_{i}$,
\begin{equation}
e^{\beta\epsilon\,{\bf v}\cdot{\bf w}}=\cosh^{N}\!\left(\beta\epsilon\right)\sum_{S}t^{\left|S\right|}\chi_{S}({\bf v})\chi_{S}({\bf w})\label{eq:sm_walsh}
\end{equation}
with $t=\tanh\beta\epsilon$ and the sum over all subsets $S$ of
the $N$ variables, so that 
\begin{equation}
M=\cosh^{N}\!\left(\beta\epsilon\right)\left[{\bf q}{\bf q}^{\top}+B\right],\quad B=\sum_{S\neq\emptyset}t^{\left|S\right|}{\bf w}_{S}{\bf w}_{S}^{\top}\label{eq:sm_B}
\end{equation}
where ${\bf w}_{S}={\bf q}\circ{\boldsymbol{\chi}}_{S}$ and $\circ$
is the elementwise product. The $S=\emptyset$ term is the same rank-one
eigenvalue as before, now multiplied by $\cosh^{N}(\beta\epsilon)Z_{\text{REM}}=e^{\beta N\zeta}Z_{\text{REM}}$,
so the domain-wall contribution appears directly as a prefactor. The
remainder $B$ is positive semidefinite of rank $2^{N}-1$, with trace
$Z_{\text{REM}}[(1+t)^{N}-1]$, but it is neither diagonal nor of
low rank: the ${\bf w}_{S}$ are not orthogonal. We have not carried
the analysis further.

\subsection{Numerical methods}

The largest eigenvalue and its eigenvector are obtained by Lanczos
iteration, with $M$ applied as a sequence of $N$ two-site operations
rather than stored, so that a matrix-vector product costs $O(N2^{N})$.
For the state-matched coupling Eq. (\ref{eq:sm_secular}) replaces
diagonalization altogether: the root is found by Newton iteration
starting from Eq. (\ref{eq:sm_y_eta}), which requires only the spectrum
$\left\{ E_{{\bf v}}\right\} $ and not the eigenvector. This reaches
$N=50$, where the spectrum is too large to store and is replaced
by a binned histogram, the bin width chosen so that the error in $p_{\text{GS}}$
is below $10^{-3}$. This is equivalent to binning the spectrum, with
the multiplicity of each bin drawn from the distribution of energies
within it rather than enumerated.

\section{The GREM chain}

\subsubsection{Setting and result}

We consider a chain of $L$ copies of the same two-level GREM, $N\to\infty$
before $L\to\infty$, with $\epsilon_{g},\epsilon_{s}>0$. 
\begin{equation}
H_{L}=\sum_{\mu=1}^{L}E_{i_{\mu}}-N\epsilon_{g}\sum_{\mu}\delta_{g_{\mu},g_{\mu+1}}-N\epsilon_{s}\sum_{\mu}\delta_{i_{\mu},i_{\mu+1}}\label{eq:H}
\end{equation}

The proof below is given for $a_{g}/\ln\alpha_{g}>a_{s}/\ln\alpha_{s}$,
called Case I below, in which the group level freezes at the higher
temperature.

We find that the chain has exactly three phases, and no other state
competes. Each level locks when its own coupling exceeds that level's
gap between the ground state and equilibrium, 
\begin{equation}
\Delta_{s,g}(T)\equiv f_{s,g}(0)-f_{s,g}(T)\label{eq:Delta}
\end{equation}
where $f_{g,s}(T)\equiv-\frac{T}{N}\ln\sum_{g,s}e^{-E_{g,s}/T}$.
The state level is unable to lock unless the group level locks with
it.

This appendix establishes the following four statements made about
the GREM chain in the main text.
\begin{enumerate}
\item Three phases occur: all copies in different groups; all in one group
but in different states; all in one state.
\item No phase with intermediate domain lengths competes.
\item The locked group is the one of lowest free energy $\mathcal{F}_{g}$.
With groups of equal size this reduces to the lowest $E_{g}$, and
that the intermediate and fully locked phases need not select the
same group.
\item The intermediate phase appears when $\alpha_{g}<\alpha_{s}$: fewer
groups than states per group.
\end{enumerate}

\subsection{Review of GREM preliminaries}

We limit ourselves here to the GREM with $n=2$ levels.

There are $2^{N}$ levels, grouped into $\alpha_{g}^{N}$ groups,
in each of them there are $\alpha_{s}^{N}$ states. Therefore, $\alpha_{g}\alpha_{s}=2$.
Let $g({\bf v})=1,..,\alpha_{g}^{N}$ be the index of the group in
which $i$ resides. The energy of a state is 
\[
E_{{\bf v}}=E_{g,g({\bf v})}+E_{s,{\bf v}}
\]
where $E_{g,g({\bf v})}$ is common to the group $g({\bf v})$ to
which $i$ belongs, and $E_{s,{\bf v}}$ independent inside the group.
The variance of $E_{k,{\bf v}}$ is $NJ^{2}a_{k}$, so that with $a_{g}+a_{s}=1$,
we have as for the REM, $\left\langle E_{{\bf v}}^{2}\right\rangle =NJ^{2}$.

\subsubsection{Density of states}

The number of groups in energy $E_{g}$ is Poisson with mean
\[
\mathcal{N}(E_{g})\sim\exp N\left(\ln\alpha_{g}-\frac{u_{g}^{2}}{J^{2}a_{g}}\right)
\]
 with $u_{g}=E_{g}/N$. This gives non-zero counts in $J^{2}\ln\alpha_{g}>\frac{u_{g}^{2}}{a_{g}}$
or $J\sqrt{a_{g}\ln\alpha_{g}}>\left|u_{g}\right|$.

For $E_{s}$ the situation is more subtle. Consider systems at the
energy ranges of width $\Delta E_{g},\Delta E_{s}$ around $\left(E_{g},E_{s}\right)$.
There are $e^{N\phi_{g}(E_{g})}\Theta\left[\phi_{g}(E_{g})\right]$
groups, and given that there are any groups at all, the mean number
of states is $e^{N\left[\phi_{g}(E_{g})+\phi_{s}(E_{s})\right]}$,
which means that the actual number of states is $e^{N\left[\phi_{g}(E_{g})+\phi_{s}(E_{s})\right]}\Theta\left[\phi_{g}(E_{g})\right]\Theta\left[\phi_{g}(E_{g})+\phi_{s}(E_{s})\right]$.
This holds \emph{even if} $\phi_{s}(E_{s})<0$, as long as $\phi_{g}(E_{g})+\phi_{s}(E_{s})>0$
(in this case this is a Poisson distribution with mean $N(...)$).
This gives the condition
\[
J^{2}\left(\ln\alpha_{g}+\ln\alpha_{s}\right)>\frac{u_{g}^{2}}{a_{g}}+\frac{u_{s}^{2}}{a_{s}}\Rightarrow J^{2}\ln2>\frac{u_{g}^{2}}{a_{g}}+\frac{u_{s}^{2}}{a_{s}}
\]

Combining the conditions, the partition function reads
\begin{align*}
Z & =\sum_{g=1..\alpha_{g}^{N}}e^{-\beta E_{g,g}}\sum_{i,g(i)=g}e^{-\beta E_{s,i}}\\
 & \sim\int_{\begin{array}{c}
u_{g}^{2}/a_{g}<J^{2}\ln\alpha_{g}\\
\frac{u_{g}^{2}}{a_{g}}+\frac{u_{s}^{2}}{a_{s}}<J^{2}\ln2
\end{array}}dE_{g}dE_{s}\mathcal{N}(E_{g})e^{-\beta E_{g}}\mathcal{N}(E_{s})e^{-\beta E_{s}}
\end{align*}
 so that
\[
f_{\text{GREM}}=-\frac{T}{N}\ln Z=-T\max_{\begin{array}{c}
u_{g}^{2}/a_{g}<J^{2}\ln\alpha_{g}\\
\frac{u_{g}^{2}}{a_{g}}+\frac{u_{s}^{2}}{a_{s}}<J^{2}\ln2
\end{array}}\left[\psi_{g}(u_{g})+\psi_{s}(u_{s})\right]
\]
 where
\[
\psi_{g,s}(u_{g,s})\equiv\ln\alpha_{g,s}-\frac{u_{g,s}^{2}}{J^{2}a_{g,s}}-\beta u_{g,s}
\]
At high $T$, both $u_{g}$ and $u_{s}$ attain their maxima in the
bulk, giving $\psi_{g,s}'(u_{g,s})=0\Rightarrow u_{g,s}=-\frac{1}{2}J^{2}a_{g,s}\beta$.
As $T$ is lowered, there are different cases according to which condition
is hit first. It is useful to define the transition temperatures
\[
T_{g,s}\equiv\frac{J}{2}\sqrt{\frac{a_{g,s}}{\ln\alpha_{g,s}}}
\]

Below, we describe one of the cases, Case I. Case II will not be covered,
as we don't use it in the system of coupled-copies below.

\subsubsection{Case I}

This scenario holds when $T_{s}<T_{g}\Leftrightarrow\frac{a_{s}}{\ln\alpha_{s}}<\frac{a_{g}}{\ln\alpha_{g}}$.

Lowering $T$, the condition $u_{g}^{2}/a_{g}<J^{2}\ln\alpha_{g}$
is saturated first at $T_{g}$, after which $u_{g}^{2}/a_{g}=J^{2}\ln\alpha_{g}$.
The second condition then becomes, using $\alpha_{g}\alpha_{s}=2$,
$\frac{u_{s}^{2}}{a_{s}}<J^{2}\ln\alpha_{s}$ which finally also saturated
at $T=T_{s}$. This means that in all three temperatures ranges, below
and above $T_{s},T_{g}$ with $T_{s}<T_{g}$, the maximization procedure
is separate for the states and the groups: in Case I, 
\[
\max_{\begin{array}{c}
u_{g}^{2}/a_{g}<J^{2}\ln\alpha_{g}\\
\frac{u_{g}^{2}}{a_{g}}+\frac{u_{s}^{2}}{a_{s}}<J^{2}\ln2
\end{array}}\left[\psi_{g}(u_{g})+\psi_{s}(u_{s})\right]=\max_{\begin{array}{c}
u_{g}^{2}/a_{g}<J^{2}\ln\alpha_{g}\end{array}}\left[\psi_{g}(u_{g})\right]+\max_{\begin{array}{c}
u_{s}^{2}/a_{s}<J^{2}\ln\alpha_{s}\end{array}}\left[\psi_{s}(u_{s})\right]
\]
 The two maximization procedures on the right-hand side is thus the
sum of two separate REM models, so the free energy reads
\begin{equation}
f_{\text{GREM}}(T)=f_{g}(T)+f_{s}(T)\label{eq:f_additive}
\end{equation}
 where
\begin{equation}
f_{g}(T)\equiv-\frac{T}{N}\ln\sum_{g}e^{-E_{g}/T},\qquad f_{s}(T)\equiv-\frac{T}{N}\ln\sum_{{\bf v}\in g}e^{-E_{s,i}/T}\label{eq:fdef}
\end{equation}
 Note also that from the maximization equation, $f_{s}$ is independent
of the group $g$. The REM-like free energies read
\begin{align}
f_{g,s}(T) & =\begin{cases}
-T\ln\alpha_{g,s}-\dfrac{J^{2}a_{g,s}}{4T}, & T>T_{g,s}\\[6pt]
-u_{0,g,s}, & T<T_{g,s}
\end{cases}\nonumber \\
u_{0,g,s} & =J\sqrt{a_{g,s}\ln\alpha_{g,s}},\quad T_{g,s}=\frac{J}{2}\sqrt{\frac{a_{g,s}}{\ln\alpha_{g,s}}}\label{eq:fexplicit}
\end{align}
 In particular $f_{g,s}(0)=-u_{0,g,s}$, and $f_{g,s}$ is concave
in $T$, being a free energy.

\subsection{Exhaustive classification of states}

Classify a state of the system $({\bf v}_{1},\dots,{\bf v}_{L})$
using a hierarchical partition of the chain into domains. First, partition
the chain into group domains, maximal runs of copies sharing a group,
of lengths $\{m_{a}\}$ with $\sum_{a}m_{a}=L$. Within each group
domain, partition into subdomains, maximal runs of copies sharing
a state, of lengths $\{l_{ab}\}$ with $\sum_{b}l_{ab}=m_{a}$. Given
the states of the entire chain this partition is well defined and
unique. Equal states imply equal groups, so a subdomain never straddles
a group domain, and every copy belongs to exactly one group domain
and one subdomain. Every state of the chain is therefore counted exactly
once.

By construction, adjacent group domains carry different groups, and
adjacent subdomains within a group domain carry different states.
When summing to obtain the partition function we neglect these constraints.
A state in which two adjacent domains happen to carry the same label
is then counted both as two domains and as one longer domain, and
the two-domain version is missing one satisfied bond, so it is suppressed
by $e^{-\beta\epsilon_{g}N}$ at the group level and by $e^{-\beta\epsilon_{s}N}$
at the state level. The correction is therefore sub-exponential provided
$e^{\beta\epsilon_{g}N}\gg1$ and $e^{\beta\epsilon_{s}N}\gg1$, as
in the $L=2$ REM calculation. To exponential accuracy the group domains
then factorize: 
\begin{equation}
Z_{L}=\prod_{a}W\left(m_{a};\{l_{ab}\}\right)\label{eq:factorize}
\end{equation}

A group domain of length $m$ containing $k$ subdomains of lengths
$\{l_{b}\}$ has $m-1$ satisfied group bonds and $\sum_{b}(l_{b}-1)=m-k$
satisfied state bonds. All $m$ copies share $E_{g}$; the copies
of subdomain $b$ share one $E_{s,{\bf v}}$, counted $l_{b}$ times:
\begin{equation}
W=e^{\beta N\left[\epsilon_{g}(m-1)+\epsilon_{s}(m-k)\right]}\sum_{g}e^{-m\beta E_{g}}\prod_{b=1}^{k}\left(\sum_{{\bf v}\in g}e^{-l_{b}\beta E_{s,i}}\right)\label{eq:W}
\end{equation}
 Using Eq. \ref{eq:fdef}, and as shown for Case I, each internal
sum is independent of $g$, and each outer sum is a level-$g$ REM
at temperature $T/m$:
\begin{equation}
\sum_{{\bf v}\in g}e^{-l\beta E_{s,i}}=e^{-l\beta Nf_{s}(T/l)},\qquad\sum_{g}e^{-m\beta E_{g}}=e^{-m\beta Nf_{g}(T/m)}\label{eq:sums}
\end{equation}
 Inserting (\ref{eq:sums}) into (\ref{eq:W}), and defining $x\equiv1/m$,
$y_{b}\equiv1/l_{b}$, and $k/m=\frac{1}{m}\sum_{b}l_{b}y_{b}$,
\begin{equation}
-\frac{T\ln W}{mN}=-\epsilon_{g}(1-x)+f_{g}(xT)+\frac{1}{m}\sum_{b}l_{b}\,h(y_{b}),\label{eq:percopy}
\end{equation}
 with $h(y)\equiv-\epsilon_{s}(1-y)+f_{s}(yT)$. The state level enters
only through a weighted average of the single function $h$, with
weights $l_{b}/m$ that are non-negative and sum to one.

\subsection{Minimization}

$1\le l_{b}\le m$ gives $y_{b}\in[x,1]$: this is the nesting constraint,
and it has no analogue in the REM. $h$ is concave, since $f_{s}(yT)$
is a free energy at a temperature linear in $y$ and $-\epsilon_{s}(1-y)$
is affine. A weighted average of a concave function is at least its
minimum over the range, so 
\begin{equation}
\frac{1}{m}\sum_{b}l_{b}\,h(y_{b})\ \ge\ \min_{y\in[x,1]}h(y)=\min\left\{ h(x),h(1)\right\} \label{eq:inner}
\end{equation}
attained with all $y_{b}$ equal. The state level therefore takes
one of two values, $y=1$ (states all different) or $y=x$ (one subdomain
spanning the whole group domain). Nothing intermediate, and no mixture
of the two.

We have thus shown that $-T\ln W/(mN)\ge G(x)$, where 
\begin{equation}
G(x)=-\epsilon_{g}(1-x)+f_{g}(xT)+\min\left\{ h(x),h(1)\right\} \label{eq:G}
\end{equation}
is concave on $x\in[0,1]$: a minimum of two concave functions, plus
an affine term. Hence $G(x)$ attains its minimum at $x\in\{0,1\}$.
The bound is attained: both minimizing choices, all $y_{b}=1$ and
all $y_{b}=x$, are realizable partitions, so $G(x)$ is the free
energy per copy and not merely a bound on it.

The possible minimizers of $G(x)$ are $x\in\{0,1\}$, with the minimum
of $h$ at either $y=x$ or $y=1$. This is four pairs $(x,y)$, but
at $x=1$ the two choices of $y$ coincide, leaving exactly three:
$(1,1)$, $(0,1)$ and $(0,0)$. Using Eqs. \ref{eq:fexplicit},\ref{eq:f_additive},
$f_{k}(0)=-u_{0,k}$, and the values of $G$ are 
\begin{align}
G(1,1) & =f_{g}(T)+f_{s}(T)=f_{\text{GREM}}(T),\nonumber \\
G(0,1) & =-\epsilon_{g}-u_{0,g}+f_{s}(T)\nonumber \\
G(0,0) & =-\epsilon_{g}-\epsilon_{s}-u_{0,g}-u_{0,s}\label{eq:three}
\end{align}
This means that there are only three expressions to compare when calculating
the free energy: all copies in different groups; all copies in the
same group but in different states; and all copies in the same state.
All other possibilities have higher free energy. Writing $\delta_{k}\equiv\epsilon_{k}-\Delta_{k}$
with $\Delta_{k}$ from (\ref{eq:Delta}), the three values are $G(1,1)$,
$G(1,1)-\delta_{g}$, $G(1,1)-\delta_{g}-\delta_{s}$, so the free
energy is 
\begin{equation}
-G=-G(1,1)+\max\left\{ 0,\ \delta_{g},\ \delta_{g}+\delta_{s}\right\} \label{eq:free}
\end{equation}

\subsubsection{Which group is selected}

The three phases are distinguished by which levels are locked. Which
group the chain settles into is a separate question, and the answer
differs between phases. In the intermediate phase the chain occupies
the group minimizing $\mathcal{F}_{g}=E_{g}+Nf_{s}(T)$. With groups
of equal size the second term is common to all of them, so the selected
group is simply the one of lowest $E_{g}$, independent of temperature:
the free-energy competition is degenerate. When the groups differ
in size, $f_{s}$ carries a group-dependent entropy and the selection
can be entirely entropic.

The two locked phases need not select the same group. The intermediate
phase picks $\arg\min_{g}E_{g}$, whereas the fully locked phase minimizes
the total energy of the deepest state in each group, $\arg\min_{g}\left[E_{g}+\xi_{g}\right]$
with $\xi_{g}$ the internal fluctuation, of order unity. Neither
depends on temperature, but they differ (for example, drawing $\xi_{g}$
from the band-edge Poisson process of the internal REM, the two coincide
only with probability of about $0.42$ at the parameters of Fig. 4(B)).
So, on crossing from the intermediate phase into the fully locked
one, the chain therefore often jumps to a different group.

\section{Antiferromagnetic coupling}

For $\epsilon<0$ the structured coupling favors neighboring copies
in opposite states, and a new phase appears in which a pair of states
alternates indefinitely along the chain, $({\bf v},{\bf v}^{\prime},{\bf v},{\bf v}^{\prime},\dots)$.
This section describes the solution, the alternatives that have and
haven't been checked, and our numerical checks.

\subsection{The alternating solution}

In the alternating state only two states appear, however long the
chain, so the entropy per copy vanishes and the free energy is purely
energetic (this is in analogy with all-equal locking for $\epsilon>0$).
Let the two carry energy densities $u$ and $u^{\prime}$ and overlap
$\alpha$. The free energy per copy per variable is 
\begin{equation}
f_{\text{alt}}=\frac{u+u^{\prime}}{2}-\epsilon\alpha\label{eq:sm_falt}
\end{equation}
This is to be minimised over the pair, subject to the pair existing.
The number of pairs with those values is $e^{Ns}$ with
\begin{equation}
s(u,u^{\prime},\alpha)=2\ln2+\phi(\alpha)-\frac{u^{2}}{J^{2}}-\frac{u^{\prime2}}{J^{2}}\label{eq:sm_s_pair}
\end{equation}
 where $\phi(\alpha)$ is defined in the main text, and counts the
states at overlap $\alpha$ from a given one. The constraint is $s\ge0$.

Minimising $f_{\text{alt}}$ subject to $s=0$ by a Lagrange multiplier
$\lambda$ gives $u=u^{\prime}=-J^{2}/4\lambda$ and $\alpha=\tanh(\epsilon/\lambda)$.
Writing $\lambda=1/\beta_{a}$, an ``inverse temperature'' belonging
to the saddle, not to the bath, gives
\begin{equation}
2\ln2-\frac{J^{2}\beta_{a}^{2}}{8}+\ln\cosh\left(\beta_{a}\epsilon\right)-\beta_{a}\epsilon\tanh\left(\beta_{a}\epsilon\right)=0\label{eq:sm_beta_a}
\end{equation}
with $u^{*}=-J^{2}\beta_{a}/4$ and $\alpha^{*}=\tanh\beta_{a}\epsilon$.
Since $\beta_{a}$ is fixed by the landscape alone, $f_{\text{alt}}=u^{*}-\epsilon\alpha^{*}$
does not depend on $T$.

This competes with the free phase, in which all copies differ, whose
free energy is
\begin{equation}
f_{\text{diff}}(T)=f_{\text{REM}}(T)-T\ln\cosh\left(\epsilon/T\right)\label{eq:sm_fdiff}
\end{equation}
the same $\zeta$ as for $\epsilon>0$. The transition is where the
two are equal. Because $f_{\text{alt}}$ is flat in $T$ while $f_{\text{diff}}$
falls with slope $-s_{\text{REM}}$, the two branches cross once and
with different slopes: the transition is first order, and the entropy
jumps by $s_{\text{REM}}$ across it. At $\epsilon=-0.5$ the crossing
lies at $T_{\text{tr}}=1.146$, above the freezing temperature $0.710$
of the all-different phase, so the alternating phase pre-empts freezing.

\subsubsection{Other cyclic solutions}

The alternating state is one member of a family. We compared the uniform
state, cycles of period $m$ states, and the all-different family,
and the alternating one wins throughout $\epsilon<0$. We have not
excluded chains that revisit states in more complicated patterns,
so this is a tested family rather than a proof, in contrast with the
$\epsilon>0$ case where the domain classification is exhaustive.

The number of cycles of length $m$ with consecutive overlap $\alpha$
and energy density $u$ throughout is $e^{Nm\left[\Omega_{m}(\alpha)-u^{2}/J^{2}\right]}$
with
\begin{equation}
\Omega_{m}(\alpha)=\ln2+\frac{1}{m}\ln\left(1+t^{m}\right)+\phi(t)+\left(t-\alpha\right)\text{atanh}\,t\label{eq:sm_Omega_m}
\end{equation}
with $t$ fixed by $\left(t^{m-1}+t\right)/\left(t^{m}+1\right)=\alpha$.
The constraint is again that there exists at least one such cycle,
$\Omega_{m}(\alpha)-u^{2}/J^{2}\ge0$, and the free energy per copy
per variable is
\begin{equation}
f_{\text{cyc}}=u-\epsilon\alpha-T\left[\Omega_{m}(\alpha)-\frac{u^{2}}{J^{2}}\right]\label{eq:sm_cycle}
\end{equation}
to be minimized over $u$ and $\alpha$. At $m=2$ this reduces to
the pair case: $\Omega_{2}(\alpha)=\ln2+\tfrac{1}{2}\phi(\alpha)$,
so $e^{2N\left[\Omega_{2}-u^{2}/J^{2}\right]}=e^{Ns}$ with $s$ of
Eq.~(\ref{eq:sm_s_pair}) at $u=u^{\prime}$. We solved this optimization
problem for general $m$, and compared the solutions numerically for
a wide range of $m$, and all values with $m>2$ have higher energies
than $m=2$.

\subsubsection{Numerics}

\begin{figure}
\begin{centering}
\includegraphics[width=0.5\columnwidth]{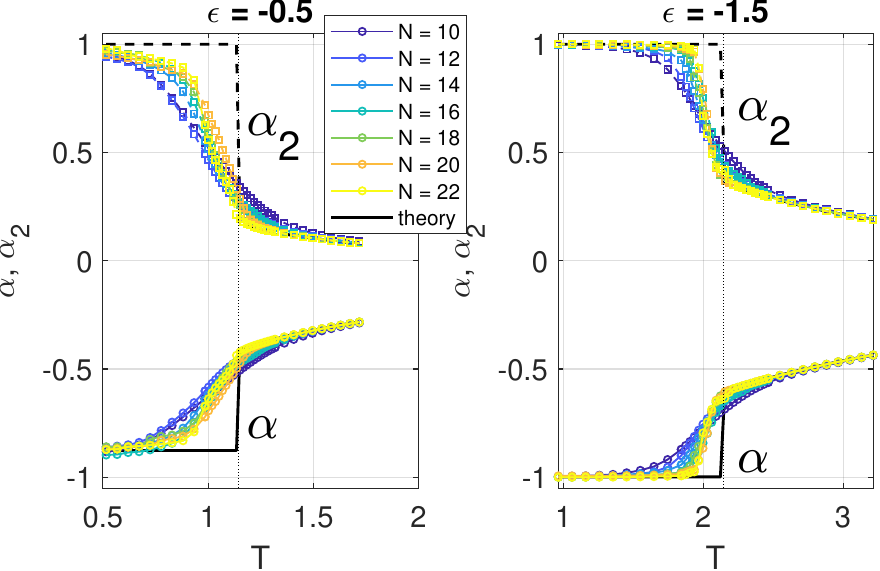}
\par\end{centering}
\caption{ The transition in structured coupling with $\epsilon<0$, transfer
matrix numerics. $\alpha$ is the overlap between nearest neighbors,
and $\alpha_{2}$ between next-nearest neighbors.}\label{fig:antiferro_transition}
\end{figure}

The transfer matrix was diagonalized for $N=10$ to $22$, measuring
the overlaps between neighboring and next-neighboring copies, see
Fig.~\ref{fig:antiferro_transition}. Two features confirm the alternating
structure: one, $\alpha_{2}$ the overlap with the next nearest neighbor,
goes to one, as it must be if the chain repeats with period two. Two,
$\alpha$ between nearest neighbors approaches $\sim-0.89$ at $\epsilon=-0.5$
and $N=22$ against the predicted $-0.876$.

The transition temperature is hard to test at these sizes. This is
the same situation as for $\epsilon>0$, with the added complication
that averaging over realizations increases the apparent width, since
each realization has its own transition temperature.

\section{Relation to the quantum REM}

A chain of coupled copies invites comparison with a Suzuki-Trotter
decomposition, and so with the transverse-field REM \citep{goldschmidtSolvableModel1990}.
The two are related at the level of the structure of the Hamiltonian,
but the QM limit requires a joint limit of large $(\epsilon,T,L)$,
which has no natural interpretation in the present model, and leads
to quite different phenomenology.

To see this, we start from the quantum model $H_{\text{QM}}=E(\hat{{\bf v}})-\Gamma\sum_{i=1}^{N}\hat{\sigma}_{i}^{x}$,
with the same $2^{N}$ energies and a transverse field $\Gamma$.
Writing its partition function at inverse temperature $\beta_{\text{QM}}$
as a path integral with $L$ slices of width $\Delta\tau=\beta_{\text{QM}}/L$,
each factor between slices is a product of single-spin matrices, and
\begin{equation}
\left\langle {\bf v}\right|e^{\Delta\tau\Gamma\sum_{i}\hat{\sigma}_{i}^{x}}\left|{\bf v}^{\prime}\right\rangle =\left(\frac{\sinh2\Delta\tau\Gamma}{2}\right)^{N/2}e^{J_{\perp}{\bf v}\cdot{\bf v}^{\prime}}\label{eq:sm_trotter}
\end{equation}
with $\tanh\left(\Delta\tau\Gamma\right)=e^{-2J_{\perp}}$. Together
\begin{equation}
Z_{\text{QM}}\propto\sum_{\{{\bf v}_{\mu}\}}\exp\left[-\Delta\tau\sum_{\mu}E({\bf v}_{\mu})+J_{\perp}\sum_{\mu}{\bf v}_{\mu}\cdot{\bf v}_{\mu+1}\right]\label{eq:sm_pathint}
\end{equation}
which is the classical chain of this paper with structured coupling:
the slices are the copies, and $L$ is their number.

Comparing with $\beta H_{L}$ in the present paper gives the dictionary
\begin{equation}
\Delta\tau=\frac{1}{T},\qquad e^{-2\epsilon/T}=\tanh\frac{\Gamma}{T},\qquad T_{\text{QM}}=\frac{T}{L}\label{eq:sm_dict}
\end{equation}
Three classical parameters therefore fix three quantum ones, each
with a different role: the classical temperature is the inverse Trotter
step, the classical coupling sets the transverse field, and the quantum
temperature is fixed by the chain length. The quantum limit is reached
taking $T_{\text{QM}}$ fixed and $L\to\infty$, which in our model
reads 
\begin{equation}
\epsilon\simeq\frac{T}{2}\ln\frac{T}{\Gamma},\ L\propto T\label{eq:sm_qrem}
\end{equation}
which gives the joint limit mentioned above.

\end{document}